\documentclass[prl,preprint]{revtex4}

\usepackage[dvips]{color}
\usepackage[dvips]{graphicx}

\begin{document}
\title{Arrow of time and non-Markovianity in the
  non equilibrium folding/unfolding of alanine decapeptide {\it in vacuo}}

\author{Simone Marsili, Piero Procacci}

\affiliation{Dipartimento di Chimica, Universit\`a di Firenze, Via
della Lastruccia 3, I-50019 Sesto Fiorentino, Italy}

\affiliation{Centro Interdipartimentale per lo Studio delle Dinamiche Complesse(CSDC),
Via Sansone 1, I-50019 Sesto Fiorentino, Italy}

\date{\today}
\begin{abstract}
\noindent
We present non equilibrium molecular dynamics experiments of the
unfolding and refolding of an alanine decapeptide {\it in vacuo}
subject to a Nos\'e thermostat. Forward (unfolding) and reverse
(refolding) work distribution are numerically calculated for various
duration times of the non equilibrium experiments. Crooks theorem is
accurately verified for all non equilibrium regimes and the time
asymmetry of the process is measured using the recently proposed
Jensen-Shannon divergence [E.H. Fend, G. Crooks {\it Phys. Rev. Lett},
  101, 090602] . Results on the alanine decapeptide are found similar
to recent experimental data on m-RNA molecule, thus evidencing the
universal character of the Jensen-Shannon divergence.  The patent non
markovianity of the process is rationalized by assuming that the
observed forward and reverse distributions can be each described by a
combination of two normal distributions satisfying the Crooks theorem,
representative of two mutually exclusive linear events.  Such bimodal
approach reproduce with surprising accuracy the observed non Markovian
work distributions.

\end{abstract}

\maketitle
 
\section{introduction}

Some time ago Crooks\cite{crooks98} derived, in the context of Monte
Carlo simulations, an exact formula involving the dissipative work of a
system driven out of equilibrium through a time dependent external
potential and in contact with a thermal bath at temperature $T=1/k_B
\beta$. This formula, ever since known as the Crooks theorem (CT),
reads:
\begin{equation}
\frac{P(x,\Lambda)}{P(\hat x ,\hat \Lambda)} = e^{\beta( W -\Delta F)}
\label{eq:crooks} 
\end{equation} 
where $P(x,\Lambda), P(\hat x ,\hat \Lambda)$ are the probabilities of
observing a forward trajectory $x$ , giving the time schedule (or
protocol) $\Lambda$, and of observing its conjugate trajectory $\hat
x$ with inverted transformation protocol $\hat\Lambda$, respectively;
$\Delta F\equiv F_B - F_A $ is the free energy difference between the
initial and final canonical ensembles and $W$ is the work done in the
{\it forward} driven non equilibrium experiment.  The Crooks formula
has been later recognized of much broader validity, and it was shown
to hold for deterministic systems in the context classical molecular dynamics
simulations\cite{evans03,procacci06,schoell06,chelli07},
Langevin dynamics\cite{kurchan99,ohkuma07}, quantum
systems\cite{tasaki00,talkner09} and verified in real\cite{collin05} and
computer\cite{park04,procacci06,chelli07c} experiments.

The essential points for Eq. \ref{eq:crooksw} to hold is that the
driven forward and reverse experiments ought to be started from 
equilibrium distributions and that the transformation
protocols of the forward and reverse process (that can involve
mechanical and thermodynamic variables\cite{chelli07} as well) are one the time reversal
of the other.  As the work done in the non equilibrium trajectory
inverts sign by time reversal, the trajectories and their
time-reversal counterpart can be labeled using the work
such that Eq. \ref{eq:crooks} can thus be also written
as
\begin{equation}
\frac{P(W|F)} {P(-W|R)} = e^{\beta( W -\Delta F)}
\label{eq:crooksw} 
\end{equation} 
where $P(W|F)$, $P(-W|R)$ are the probability of observing a work $W$
  in the forward and reverse experiment. Eq. \ref{eq:crooksw} says
  that trajectories that are highly dissipative (i.e. $W-\Delta F >> 0
  $) in the forward sense are difficult to observe in the reverse
  sense since for such trajectories the dissipation of
  its time-reversal counterpart would be negative, thus transiently
  violating the second law.  In the functional form of Eq. \ref{eq:crooksw}, the
  Crooks theorems applies, with some provisions\cite{hummer01} related
  to the form of the external driving agent, to the controlled
  mechanical manipulation of a single molecule through optical
  tweezers\cite{collin05} or atomic force microscopy.\cite{imparato08} We conclude
  this introductory remarks by stating that Eq. \ref{eq:crooksw}, 
  one of the very few {\it exact} equations in non equilibrium
  thermodynamics, holds for any regime: for instantaneous pulling we
  have that $W=H_B - H_A$ and, by averaging over all trajectories, one recovers the
  Zwanzig\cite{zwanzig54} formula $<e^{-\beta(H_B-H_A)}>_A = e^{\beta
  \Delta F}$. For infinitely slow pulling, i.e for quasi-static
  reversible transformations, $W=\Delta F$ and the forward and
  backward distribution are indistinguishable and
  $P(W|F)=P(-W|R)=\delta(W-\Delta F)$.

Recently there has been considerable progress in the interpretation of
non equilibrium experiments coming both from measurements on single
molecules using AFM or optical traps\cite{feng08} and from
deterministic or stochastic simulations.\cite{kawai07} Feng and Crooks
proposed to use the Jensen-Shannon divergence\cite{lin91,cover06} (JSD) between the probability
of a trajectory and its time-reversal conjugate as a definition and a
measure of the {\it time asymmetry} in a thermodynamics system.  If we use
the work $W$ (which changes sign by time reversal) as a label for
trajectories, then JSD can be written in terms of work
distributions as
\begin{eqnarray}
{\rm JSD} & = & \frac{1}{2} 
\int P(W|F) \ln \frac{2 P(W|F)}{P(W|F) + P(-W|R)}  dW \nonumber \\ 
& +  &  
\frac{1}{2}  \int P(-W|R) \ln \frac{2 P(-W|R)}{P(W|F) + P(-W|R)}  dW.
\label{eq:js}
\nonumber
\end{eqnarray}
$\rm JSD$ can be shown\cite{feng08} to be equal to the average gain of
information about the orientation of time's arrow from one single
realization of the experiment.  This quantity, plotted against the
average dissipation obtained in the forward and reverse driven
experiments, goes to zero for reversible processes, and to one full nats
of information $\ln 2$ (i.e. 1 bit) when the two distributions do not
overlap (i.e. for large average dissipation).  In this latter case, it
is easy to assign an observed trajectory (taken from the pool of
forward and reverse non equilibrium experiments) to one of two
distributions, or, stated in other words, it is easy to guess, from
the analysis of one single random trajectory, in which direction the
time is flowing.  On such basis, when plotted against the average mean
dissipation, JSD may then give indication on the
energetic cost (i.e. the dissipation needed) to ensure that a
molecular process (e.g. a molecular motor) advances in time.  For
Markovian (linear) systems, the work distributions are always
Gaussian\cite{park04,procacci06} with variance twice the average
dissipation. In this case, JSD {\it vs} dissipation is analytic and
identical for all Markovian system.  Therefore, Eq. \ref{eq:js} can
also be used a measure of the non linearity of the system.

In the context of non equilibrium thermodynamics, similar concepts
were put forward recently by Kawai, Parrando and Van den
Broeck.\cite{kawai07} For system perturbed far from equilibrium through
driven forward and time reversal protocols, they derived a remarkable
exact formula connecting the relative entropy of the two conjugate
phase space density of system measured at the same but otherwise
arbitrary point in time to the average dissipation in the forward
experiment.  Exploiting the fact that the system is deterministic and
that the work inverts its sign by time reversal, and labeling each
phase space point in terms of (future) works, the Kawai-Parrondo-Van
der Broeck formula can be straightforwardly written in terms of work
distribution alone as
\begin{eqnarray}
<W>-\Delta F  & =  & k_BT D((P(W|F)|P(-W|R))  \nonumber \\
& = & k_B T  \int dW P(W|F) \log \left ( \frac{ P(W|F)}{P(-W|R)}
  \right   ) 
\label{eq:kl} 
\end{eqnarray}
Eq. \ref{eq:kl} can be easily derived form the Crooks theorem
Eq. \ref{eq:crooksw}.  The integral on the {\it lhs} is the
Kullback-Leibler divergence (KLD),\cite{cover06} a strictly positive
quantity measuring, in information theory, the expected extra
message-length per datum that must be communicated if a code that is
optimal for a given (wrong) distribution $P(-W|R)$ is used, compared
to using a code based on the (true) distribution $P(W|F)$. In general
the KLD is not symmetric, i.e. if $q,p$ are two non identical
distributions, $D(p|q) \ne D(q|p)$. For Markovian systems,
however, the KLD is always symmetric. Moreover, for such systems, $k_B T$
times the KL divergence can be calculated analytically 
yielding the dissipation $\beta \sigma^2/2$, with $\sigma^2$ being the
variance of the Gaussian distribution.  KL between forward and reverse
distributions has the same characteristics of the JSD divergence,
being the former like the latter both a measure of the time asymmetry
(i.e. of the possibility for distinguish in which sense the time is
flowing ) and of non linearity. However KL, as suggested by Kawai {\it
  et al.}, could be effectively used as a tool for obtaining a better
upper bound of the free energy than the average work $W$.  This is so
since, according to the chain rule,\cite{cover06} the relative
entropy (or KL divergence) decreases upon coarse graining. An
extremely simple coarse graining scheme could be that of approximating
coarse grained histograms of the forward and backward work
distribution with the best linear model satisfying the Crooks theorem.
This approach has been advocated recently by Forney {\it et al}.
\cite{forney08} in the context of steered molecular dynamics of
decaalanine {\it vacuo} along the end-to-end distance.  These authors,
in their so-called {\it FR} method\cite{kosztin06,forney08}, produce a coarse
grained histogram with few work measurements in both directions
that are then fitted using a linear (Markovian) model.  However, when
the driven coordinates exhibit clear non linear effects (i.e. the noise due
to all other ``solvent'' coordinates is not white or Gaussian), as is the case of
folding and refolding of small proteins along the end-to-end distance,
then other less simplistic coarse grain schemes could and should be
adopted. 

In this paper we further develop the concepts of time asymmetry and
coarse graining introduced in Refs. \onlinecite{kawai07,feng08} by presenting extensive
non equilibrium molecular dynamics simulation data of unfolding and
refolding process of decaalanine {\it in vacuo} performed with the
deterministic Nos\'e-Hoover thermostat at 300 K.  In spite of the fact
that decaalanine {\it in vacuo} has been extensively studied in the
recent past by non equilibrium computational techniques\cite{park04,procacci06,forney08},
the rationalization and interpretation of the observed data is still a
matter of debate. $\alpha$-helix formation is also important {\it per se}
and as a paradigm for an elementary folding/unfolding process.  

Our results on decaalanine are interpreted by means of the JSD and KLD
quantities above introduced. We further present a simple coarse grain
and totally general model satisfying the CT based on the assumption of
the occurrence, in the refolding process, of two mutually exclusive
events. Such simple coarse grain dual model explains many features of
the observed work distributions and can be rationalized with the
existence of two competing minima for low values RC in decaalanine,
i.e.  one of enthalpic nature (the helix), easily accessible, in the
refolding process, at low dissipation regimes, and the other of
entropic origin corresponding to a manifold or misfolded coil
structures which emerges at large dissipation when trying to rapidly
refold decaalanine from extended structures.  This view appears to be
quite general and is fully consistent with the rugged funnel picture
of the folding process, in the sense that escaping the rugged funnel
from below is a much tamer process than reentering the funnel from
above. 

The present paper is organized as follows.  Sec. II is
dedicated to the description of the systems and of the methods used in
the non equilibrium simulations.  In sec. III we present the computer
experiment results of the unfolding/refolding process of a single
molecule of decaalanine along with a discussion focusing on the
thermodynamic and microscopic aspects and on their rationalization in
terms of a coarse grain description of a systems of
general validity in the protein space.  Conclusive remarks and futures
perspective regarding the applicability of the presented methodology
to real experiments are presented in Sec. IV.

\section{Methods}  
In this section we provide the technical details on the steered
molecular dynamics simulations of the Alanine deca-peptide (A$_{10}$)
{\it in vacuo}. The unperturbed system is described with the all-atom
force field CHARMM whose parameters are given in
Ref. \onlinecite{mackerrell98}.  A constant
temperature of 300 K is imposed through a Nos\'e-Hoover thermostat.\cite{nose84}
The resulting deterministic equations of motions are efficiently
integrated using a reference system propagator alogorithm
\cite{marchi98} at three time steps, 3.0 fs for medium and long
range non bonded interactions (no cut-off is imposed ), 1.5 fs for
torsional potential involving hydrogen atoms and for short-ranged (14)
non-bonded interactions, and 0.5 fs for stretching and bending
potentials.   The non equilibrium computer experiments from a
folded ($\alpha$-helical) to an extended ({\it all trans}) structure (called
{\it forward} process) and {\it viceversa} are done according to the
following scheme proposed by Park and Schulten\cite{park04}. The N atom of
the N-terminus residue is constrained to a fixed position and attached
to the N atom of the C-terminus, though a stiff harmonic spring
(i.e. by adding a stretching potential to the unperturbed Hamiltonian)
of adjustable equilibrium distance of the form
\begin{equation}
V(t) = \frac{k}{2} \left[\zeta - \zeta(t)]
\right]^2
\label{eq:guidpot}
\end{equation}
with $\zeta(t)$ being the time-adjustable equilibrium distance
allowing the system to move along the $\zeta$ (end-to-end distance )
coordinate. The driven unfolding (and refolding) of A$_{10}$ along
$\zeta$ is bound to occur along the $\alpha-$helix axis, by means of a
bending constraint imposing the N atom of the N-terminus, the N atom
of the C-terminus and a distant dummy atom at fixed position to all
lie on the axis of the helix.  The force constant $k$ of the external
potential used for guiding the processes (Eq. \ref{eq:guidpot}) is 400
kcal mol$^{-1}$ \AA$^{-2}$. Such a large value is used to minimize the
possible negative impact of the stiff spring
approximation\cite{park04} in the calculation of the free energy
between the initial and final state.  The conjugated time protocols $\Lambda$
and $\hat \Lambda$ for the forward and reverse non equilibrium
experiments are defined by setting in Eq. \ref{eq:guidpot} the
corresponding time dependent equilibrium distances $z(t)$ and $z(\hat
t)$ as
\begin{eqnarray}
z(t) &  = & \zeta_i + ( \zeta_f - \zeta_i ) \frac{t}{\tau} = \zeta_i +
v(\tau) t \nonumber  \\
z(\hat t ) & = & \zeta_f + ( \zeta_i- \zeta_f ) \frac{t}{\tau} =
\zeta_f - v(\tau)t 
\label{eq:timeprot} 
\end{eqnarray}
where $\zeta_i$ and $\zeta_f$ are the initial and final values of the
reaction coordinate, $\tau$ is the total (simulation) time of the non
equilibrium experiment and $v(\tau) = \pm (\zeta_f -\zeta_i)/\tau $ is
the (constant) pulling speed.  In the present study, according to
previous works\cite{park04,procacci06}, we set $\zeta_i = 15.5$
\AA~  and $\zeta_f = 31.5$~\AA.  The sampling at fixed value of $\zeta$
is achieved again by using the potential of Eq. \ref{eq:guidpot} with
$\zeta(t)=\zeta_i=15.5$~\AA~ for constraining the system at the
end-to-end distance of the $\alpha$-helix and with
$\zeta(t)=\zeta_i=31.5$~\AA~ for constraining the system at the
end-to-end distance of the all trans extended structure.  In this
manner, trough ordinary equilibrium simulations in the canonical
ensemble, we sampled, by saving the configuration at regular intervals
of 2 ps, 504 initial phase-space points for the $\alpha$-helix state
and 504 initial phase-space point for the extended state. Starting
form these points, we then did forward and reverse non equilibrium
molecular dynamics experiments applying the time dependent potential
of Eq. \ref{eq:guidpot} for various time protocols (i.e. at various
pulling constant speed, corresponding to the duration $\tau$ values of
ranging from 0.021 to 4.2 ns). In particular, for {\it each} time
protocol we did 504 forward and 504 reverse non equilibrium
experiments for a total simulation time of 11.15 $\mu$s . All non
equilibrium simulations were done in parallel on a 32 node
Intel CPU X9650  cluster using an in-house parallel version of the
program orac\cite{procacci97}.  The work done on A$_{10}$ in each of the
non equilibrium experiments is calculated through
\begin{equation}  
W  =  \int_0^{\tau} K (\zeta - \zeta(t) ) v(\tau)  dt
\label{eq:workj}
\end{equation} 

\section{Results and  Discussion}

\begin{figure}
\begin{center}
\caption{} 
\label{fig1}
\includegraphics[scale=1.0,clip]{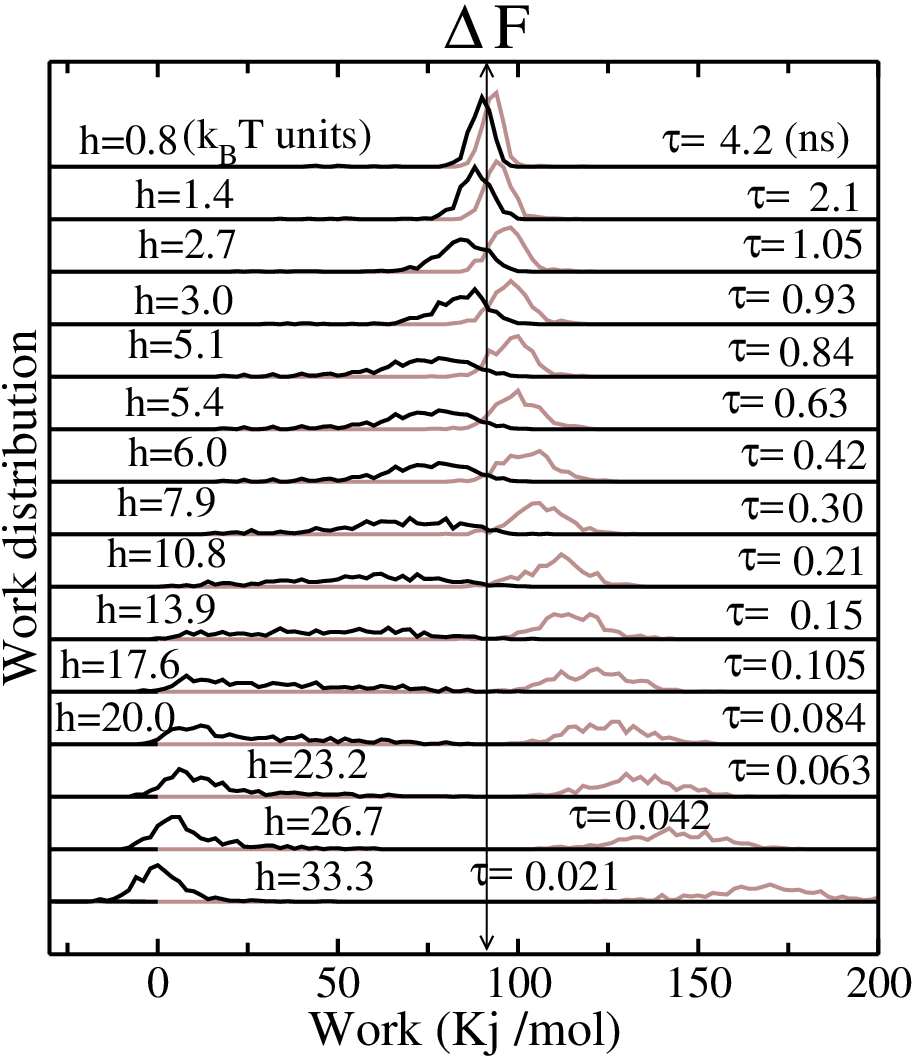}
\end{center}
\end{figure}

In Fig \ref{fig1} we show the forward (unfolding) $P(W|F)$ and
backward (refolding) $(P(-W|R)$ work distribution obtained with the
various time protocols by means of the computational methods described
in the previous section.  As expected, the two conjugated work
distributions approach to each other the longer the duration of the
non equilibrium experiment, i.e. the more reversibly is done the
transformation.  The conjugated work distributions appear to meet
approximately at the same value of $W=\Delta F$ no matter what time
protocol is used, in full agreement with the Crooks theorem
\ref{eq:crooksw}.  The free energy difference $\Delta F$ between the
helix ($\zeta=15.5$\AA) and the extended ($\zeta=31.5$~\AA) structure
can be estimated with rather good accuracy using the Bennett acceptance
ratio\cite{shirts03,bennett76} formula already starting from
$\tau=0.105$ ns, where the two work distribution overlap
significantly. Using the Bennett formula, we consistently obtain
values between 92 and 94 kJ/mol, with an average value of $\Delta
F=93.3 \pm 0.5 kJ/mol^{-1}$ in full agreement with previous estimate
of the unfolding free energy of
decaalanine\cite{park04,procacci06}. Detailed data regarding $\Delta
F$, dissipated work and variance of the distributions are reported in
Table \ref{table1}. From Inspection of a Figure \ref{fig1} and from
the data of Table \ref{table1} we see that, while the forward
distribution $P(W|F )$ preserves an approximately Gaussian shape for
all time protocols, the reverse distribution show a markedly non
Gaussian shape at all times.  In particular, the reverse distributions
are characterized by a long tail that, for $\tau< 0.1$ ns and $\tau >
0.2$ ns, lies on the right and of the left of the maximum of the
distribution, respectively. As we shall see later in the discussion,
this peculiar behaviour of the non equilibrium refolding of A$_{10}$
is a signature of competitive mutually exclusive events, i.e. the
formation of the $\alpha$-helix (for $W>60$ kJ mol$^{-1}$, i.e. at low
dissipation) form one hand and the evolution towards {\it misfolded} structures
(for $W < 25$ kJ mol$^{-1}$, i.e. at high dissipation) form the other hand.

\begin{figure}
\begin{center}
\caption{} 
\label{fig2}
\includegraphics[scale=0.6,clip]{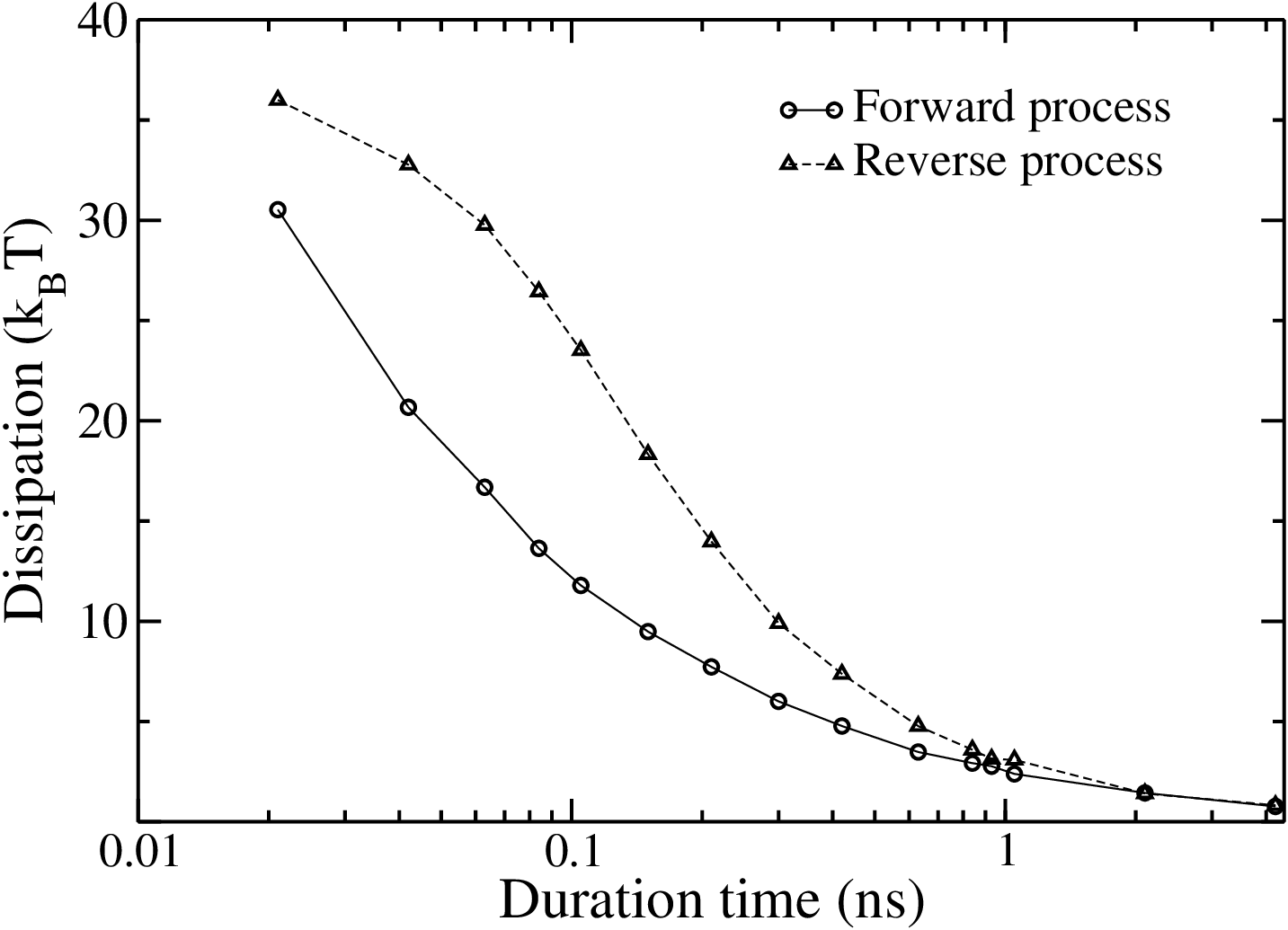}
\end{center}
\end{figure}

The asymmetry in the behaviour of $P(W|F)$ and $P(W|R)$ distribution
in $A_{10}$ is shown in Fig. \ref{fig2} where we report the forward
and reverse dissipation (see also Table I) against the duration time
$\tau$ of the non equilibrium forward and reverse experiments.  The
reverse process is consistently more dissipative than the forward for
all duration time.  Beyond $1$ ns, the dissipation for the reverse and
forward processes becomes identical and small compared to $\Delta F$,
indicating that the non equilibrium experiments are performed in
conditions of {\it quasi}-reversibility.  Such behaviour of the
dissipation of the refolding process {\it vs} the duration time of the
non equilibrium experiment could have been easily guessed directly
form Fig. \ref{fig1} by following the trends of the maxima of the
distributions as function of the duration time of the experiments.
The ``transition time'' between reversible and {\it quasi}-reversible
regimes (approximately falling between 0.8 and 1.5 ns) must be
ultimately connected to either the non equilibrium time protocol or to
some inherent structural property (e.g. potential of mean force along
$\zeta$) and dynamical property (e.g. friction and diffusion
coefficients along $\zeta$) of the system under investigation or,
again, to both.

\begin{figure}
\begin{center}
\caption{} 
\label{fig3}
\includegraphics[scale=0.6,clip]{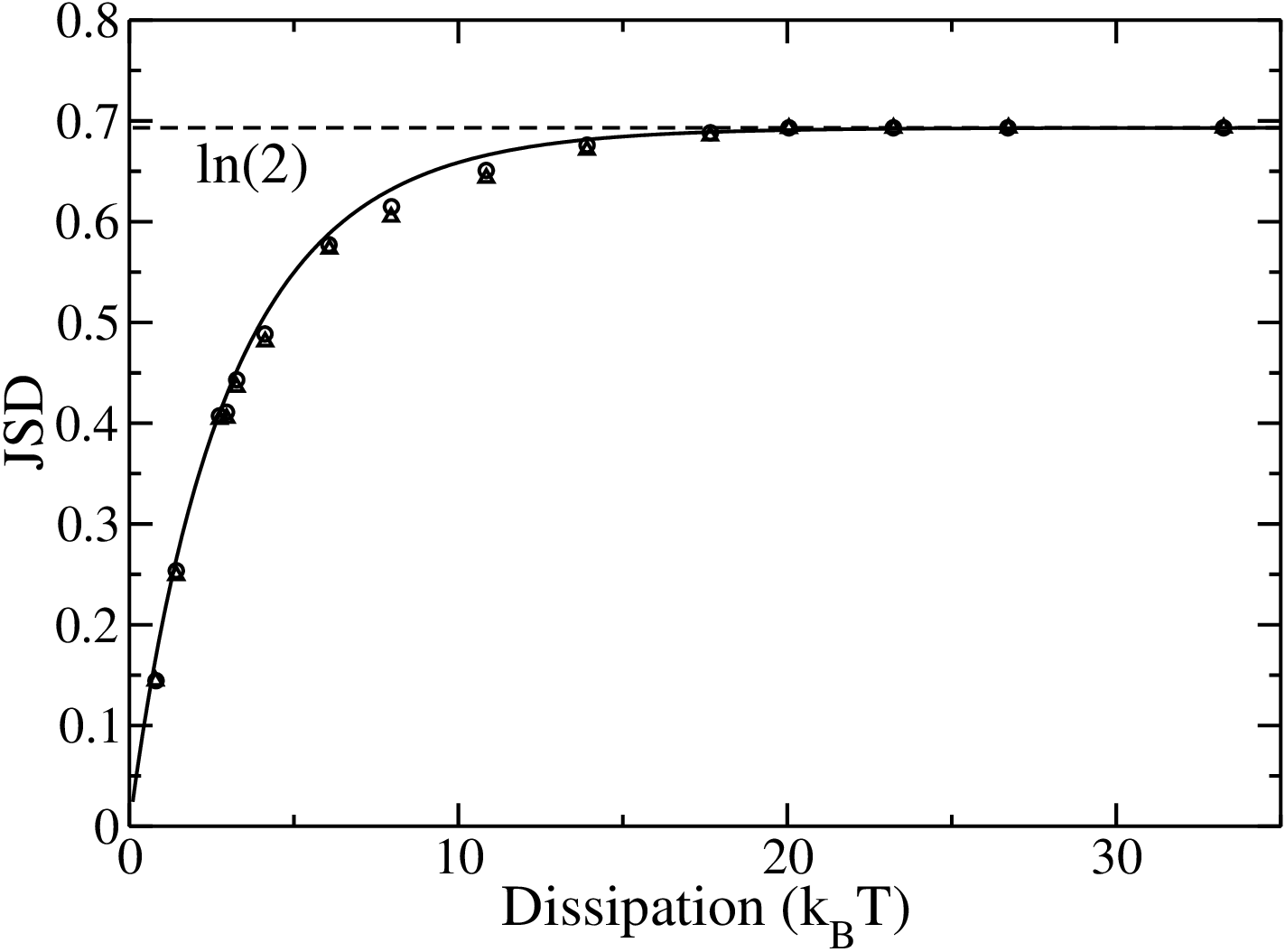}
\end{center}
\end{figure}

In order to assess the time-asymmetry and non linearity of A$_{10}$, the
data are used to compute the Jensen-Shannon divergence as a function
of the average dissipation ${\cal D} = \frac{1}{2}(<W>_f + <W>_r) $ in
$k_b T$ units.  To this aim we use directly Eq. \ref{eq:js} which can
be applied to the work data without any a prior knowledge of $\Delta
F$.  The results (triangle symbol) are reported in Fig. \ref{fig3}.
The solid line correspond to the universal Markov model where $P(W|F)$
and $P(-W|R)$ are normal distributions with equal variance and with
variance and dissipation related by $<W> - \Delta F = \beta
\frac{\sigma^2}{2}$.  Expectedly, the JSD follows closely the Markov
model for average dissipation below 4 $k_B T$, i.e. when the process
is quasi-reversible and above 13 $k_B T$ i.e. when the two
distributions have negligible overlap and the time asymmetry
approaches is limiting values of $\ln 2$ nats.  Remarkably, this
limiting value of the JSD is reached at an average dissipation that is
close to the corresponding dissipation that can be
extrapolated from the experimental data on the unfolding/refolding of
RNA molecule reported recently by Feng and Crooks (see Fig. 2 of
Ref. \onlinecite{feng08}). This suggests a universal behaviour of the JSD in real
systems, thereby strengthening the idea \cite{feng08} that the encoding
cost to ensure that a molecular process advances in time amounts to
few $k_B T$, being rather insensitive to specificity of the molecular
process itself.

In spite of the non linearity (see Table
\ref{table1} and Fig. \ref{fig2}) of A$_{10}$ along $\zeta$, especially evident at
intermediate dissipation regimes (i.e for $ 0.1 < \tau < 0.5$ ns ),
the Jensen-Shannon divergence {\it vs} dissipation appears quite
insensitive to such non linearity being nearly indistinguishable
from the universal Markov Jensen-Shannon divergence (solid line) for
all dissipation energies.  We must stress here that the error in the
Jensen-Shannon divergence {\it vs} dissipation is small (with error
bars of the order of the height of the triangle symbols in
Fig. \ref{fig3}), reflecting the small error in the determination of
$\Delta F$ itself. In order to show this more quantitatively, we have
also calculated the JSD using the alternative Eq. 7 of
Ref. \onlinecite{feng08} which requires a prior knowledge of $\Delta
F$ (through, e.g., the Bennett's method).  As one can see form
Fig. \ref{fig3},  the JSD
calculated with this method (circle symbol) follows closely that of
the direct method Eq. \ref{eq:js}.  The insensitivity of JSD {\it vs} the
mean dissipation to the non linearity of the system is probably due to
the fact that the JSD itself is a symmetries average of the two
Kullback divergence between the forward and reverse distribution with
respect to the {\it average} of the two distributions. 

The non linearity of the folding/unfolding process of A$_{10}$ for
$\tau < 1$ ns does not allows the use of a the simple Markov approach
to satisfactorily reproduce, in this time range, the observed
distribution and at the same time satisfy the CT.  As an example of
such inability, in Fig. \ref{fig4} we show the best Markov model
fitting the data for $\tau=0.105$, $\tau=0.21$ ns and at the same time
satisfying the Crooks theorem with $\Delta F=93.3$ KJ mol$^{-1}$.  The
inadequacy of the Markov model is not surprising since the driven
end-to-end distance is not a ``good'' coordinate, i.e. the modulations
of remaining (``solvent'') coordinates on $\zeta$ do not produce a
white noise. The memory effects in $\zeta$ (see e.g Fig \ref{fig2})
indicate that there must be some other important orthogonal coordinate
besides $\zeta$ that should be included in the model. We stress here
that the pure Markov model is a coarse graining of the information
regarding the microscopic detail of process in the sense that one
attempts to describes the the full information given by the
experimental histograms of the forward and reverse work with two CT
related Gaussian.  This elementary coarse-graining is the so-called
{\it FR} model\cite{kosztin06}.
   
In an effort to go beyond the simple Markov model or {\it FR} model, following Feng
and Crooks\cite{feng08}, we now assume that the forward and reverse
true distributions $P(W|F)$, $P(-W|R)$ can be approximated by 
the distributions ${\cal P}(W|F)$, ${\cal P}(-W|R)$, each given by
a linear combination of {\it two} normal distributions. i.e.
\begin{eqnarray}
{\cal P}(W|F) & = & p{\cal N}(w_1,\sigma_1 ) + (1-p){\cal N}(w_2,\sigma_2 )
\nonumber \\
{\cal P}(-W|R) & = & q{\cal N}(w_1 - \beta\sigma_1^2,\sigma_1 ) + (1-q){\cal N}(w_2-\beta\sigma_2^2,\sigma_2 ) 
\label{eq:dual}
\end{eqnarray}
where ${\cal N}(w,\sigma )$ is a normal distribution with mean $w$ and
variance $\sigma$ and $0\le p \le 1$. The form of ${\cal P}(-W|R)$ is
a trivial consequence of the CT, Eq. \ref{eq:crooksw}.
Eq. \ref{eq:dual} implies that the forward non equilibrium process is
described by two mutually exclusive events occurring with probability
$p$ and $(1-p)$ in the forward process and $q$ and $1-q$ in the
reverse process with mean dissipation satisfying the Crooks theorem.
We stress here that also such bimodal scheme is a coarse graining of
the full available microscopic information provided by the experimental
histogram $P(W)$'s.  The model can be of course complicated by
combining an arbitrary number of normal distributions allowing for many
competing events.  However, we shall see in the forthcoming discussion
that the simple bimodal scheme, Eq. \ref{eq:dual}, captures the
essential features of the ``experimental'' distributions based on the
full microscopic information.
 
Going back to Eq. \ref{eq:dual}, the probabilities $p$ and $q$ are not
free parameters as the the CT and the normalization condition set a
twofold constraint on the coefficients of the combinations
In fact, by using the Crook theorem,
Eq. \ref{eq:crooksw}, in Eq.  \ref{eq:dual}, the condition of
normalization on the probability densities $P(W|F)$ and $P(-W|R)$ requires that
\begin{eqnarray} 
p  & = & \frac{1 - e^{\beta(\Delta F-w_2 + \frac{1}{2} \beta\sigma_2^2)}} 
{ e^{\beta(\Delta F-w_1 + \frac{1}{2} \beta\sigma_1^2)}  -
  e^{\beta(\Delta F-w_2 + \frac{1}{2} \beta\sigma_2^2)} }  \\
\nonumber 
q & = & \frac{(1 - e^{\beta(\Delta F-w_2 + \frac{1}{2}
    \beta\sigma_2^2)}) e^{\beta(\Delta F-w_1 + \frac{1}{2} \beta\sigma_1^2)}}
{ e^{\beta(\Delta F-w_1 + \frac{1}{2} \beta\sigma_1^2)}  -
  e^{\beta(\Delta F-w_2 + \frac{1}{2} \beta\sigma_2^2)} }
\label{eq:ctbimod} 
\end{eqnarray}

Since $p$ and $q$ are probabilities, not all the values of the free
parameters $w_1,\sigma_1,w_2,\sigma_2$ are allowed.  We now define 
the variables $x = \Delta F-w_1 + \frac{1}{2} \beta\sigma_1^2$ and 
$y = \Delta F-w_2 + \frac{1}{2} \beta\sigma_2^2$.   
In Fig. \ref{fig5} we plot the functions $p(x,y)$ and $q(x,y)$ on the domain of the
variables $x = \Delta F-w_1 + \frac{1}{2} \beta\sigma_1^2$ and $y =
\Delta F-w_2 + \frac{1}{2} \beta\sigma_2^2$, for which $0\le p< \le 1
$ and $0\le q\le 1 $.  As it is well known, the allowed values for the variance and
the mean in a pure Markov model obeying the Crooks theorem stays on
the line $w=\Delta F + \beta \sigma^2/2$. Analogously,  Figure
\ref{fig5} represents the two dimensional domain set by the CT theorem for a bimodal Markov
model. When $p=1$, then $e^{\beta(\Delta F-w_1 +
  \frac{1}{2}\beta\sigma_1 )}=1$ such
  that $ \Delta F - w_1 = \frac{1}{2} \beta\sigma_1^2$ and $q=1$, thus
  recovering the single Gaussian Markov model.

We now adopt the model based on the coarse grain bimodal
representation Eq. \ref{eq:dual}, in order to reproduce the true work
distributions. In Table \ref{table2} we report the parameters obtained
from the fit using the bimodal distributions as a function of the
duration time.  The loss of information due to coarse graining with
respect to the true (measured) distribution is measured by the KL
divergence (Eq. \ref{eq:kl}) between the mean of the true distributions
$P(W|F) + P(-W|R)$ and the mean of the (absolutely continuous) coarse
grain distribution ${\cal P}(W|F) + {\cal P}(-W|R)$. Large values of
KL means great loss of information in the coarse graining.

We see that the bimodal approach, Eq. \ref{eq:dual} has consistently
smaller KL's with respect to the purely Markov model (shown in
\ref{table3}) at all times. A visual example of the surprising
accuracy of Eq. \ref{eq:dual} in reproducing the true distributions is
shown in Fig. \ref{fig5} where the true distributions and the bimodal
distribution of Eq. \ref{eq:dual} are compared for various short and
intermediate duration times.  By inspection of Table \ref{table3}, one
can see that the relative probability of the two mutually exclusive
event underlying the reverse distribution depends on the rate with
which the non equilibrium experiment is done. At short duration times
($\tau=0.021$ ns), the highly dissipative events in the refolding of
Alanine decapeptide are overwhelmingly more likely than the non
dissipative event, while {\it most} of the ``refolding''
trajectories produce a misfolded structure with $\zeta = 15.5$
\AA. When the rate of the non equilibrium experiment is slower
(e.g. at $\tau=0.21$ of $\tau = 0.3 $ ns), then the two competing
event (misfolding {\it vs} folding) becomes of comparable probability.
Expectedly, in the unfolding process for all duration duration times,
the dissipative event has consistently a much larger probability than
the non dissipative event (see Table \ref{table3}). In fact, while on
one hand the misfolding of $A_{10}$ starting from an extended
structure is a probable outcome in a fast refolding process, on the
other hand in the non equilibrium unfolding process it is not so
likely to disrupt the helix doing {\it less} work than the needed reversible
work.  Remarkably, the CT automatically balances these mutually
dependent probabilities $p$ and $q$ through Eq. \ref{eq:ctbimod}.

The results that we have presented show that a coarse grain scheme
based on only two mutually exclusive linear event, yielding a work
distribution that is a combination of two Gaussian distributions with
linear coefficients satisfying the Crooks theorem, explains the
observed non linear work distributions at short and intermediate times
with surprisingly good accuracy. The success of the bimodal approach
in reproducing the essential features of the true distributions at
short and intermediate duration times of the non equilibrium
experiments allows to sketch out an elementary microscopic picture: in the forward direction
only one path is possible and the process is approximately
Markovian. In the reverse direction (refolding), several other
metastable minima at $\zeta=15.5$ \AA~ can be visited depending on the
dissipation (i.e. on the duration time of the experiment).  At very
fast rate, virtually no hydrogen bond has to time to form and a
misfolded coil structure is systematically formed.  At intermediate
rates more paths are possible towards variously misfolded structure
(included distorted helices) with a probability balance between these
paths that depends on the the duration time (i.e on the mean
dissipation): the slower the process, the smaller the dissipation, the
larger is the fraction of refolding trajectories producing the helix.
At duration time between 0.6-2 ns, the refolding process ends up {\it
  mostly} in the helical structure.  For duration time beyond 3 ns,
refolding is virtually non dissipative (reversible) and only the
$\alpha$-helix minimum is visited.  The existence of the misfolded
minima at $\zeta=15.5$ \AA, that have a negligible probability at the
canonical equilibrium, emerges in the refolding process in the fast
pulling/large dissipation regime.  The rare event in these dissipative
regimes is the correct folding of decaalanine to the enthalpic minimum.
As the regime becomes less and less dissipative (i.e more reversible),
the rare event becomes the formation of misfolded coils.  The
dissipation in the refolding process, that can be simply modulated by
varying the duration time of the non equilibrium experiment, may be
thus a mean to ``see'' minima in the folded (or native) structure that
are hard to detect at equilibrium.

\section{Conclusion}  
In this paper we have studied the Alanine decapeptide in vacuo at 300
K and analyzed its behaviour in driven out of equilibrium classical
molecular dynamics simulations. Applying an external potential, we
produced classical trajectories starting form the $\alpha$-helix
structure and ending to a fully extended {\it all trans} structure and
{\it viceversa}.  The bidirectional non equilibrium experiments were
done at various pulling rate, with duration time ranging from 0.021 ps
to 4.2 ns. For each bidirectional experiment at a given pulling rate
we calculate the forward and reverse work distribution and apply the
Bennett acceptance ratio to estimate the free energy difference
between the folded and unfolded state, thus evaluating the dissipative
work spent during the non equilibrium processes.  We found that the
folding/refolding process is markedly non Markovian for
duration time $t< 1 $ 1-1.5 ns and that in such pulling rate regime
the reverse process consistently dissipates more than the forward
counterpart.  For duration time $\tau> 2$ ns, the system becomes
reversible, exhibiting equal forward and reverse mean dissipation
$W_{d}=\beta \sigma^2/2$ with the $\sigma^2$ being the variance of two
identical normal distributions. Using our non equilibrium trajectories
of A$_{10}$ and the corresponding work distributions, we have measured
the Jensen-Shannon divergence as a function of the mean forward and
reverse dissipation.  This quantity is a convenient metric for the
``irreversibility'' of the system, i.e. for the ability, given a
pulling regime yielding a given mean dissipation, to figure out in
which direction time is flowing from one random realization of the
experiment.  Remarkably, the behaviour of the Jensen-Shannon
divergence for the Alanine decapeptide {\it in vacuo} closely follows
that observed in a recent single RNA molecule experiment,\cite{collin05}
thereby strengthening the recently proposed idea \cite{feng08} that the
encoding cost to ensure that a molecular process advances in time is
independent of the system and amounts to 4-10 $k_bT$.  In the case of
the Alanine decapeptide, which shows a strongly non linear behaviour,
the Jensen-Shannon divergence plotted against the dissipation has been
nonetheless found to approximately follow the JSD for a purely Markov
model. Such surprising insensitivity of the JDS {\it vs} dissipation
to non linearity is yet another confirmation of its universal
character.

The observed forward and reverse work distributions in $A_{10}$ cannot
be fitted satisfactorily for fast and intermediate pulling speed with
normal distributions satisfying the Crooks theorem, thereby reflecting
the fact that the process in such regimes is non Markovian (i.e. the end-to-end
coordinate exhibits memory effects).  Alanine decapeptide behaves
linearly only for sufficiently slow pulling rates ($\tau \ge  1 $  ns ).   
Following a suggestion by Feng and Crooks\cite{feng08}, we thus fitted the
observed distribution using combination of two normal
distributions. This approach implies that both the forward and the
reverse process can be described by two rather than one solvent
modulated processes, whose relative probability (i.e. the ratio of the
linear coefficients of the combination) is regulated by the Crooks
theorem and by the pulling rate of the non equilibrium experiment.  We
found that such a simple model can reproduce with surprising accuracy the
observed distributions at short and intermediate pulling rate.  At
short rates the reverse distribution has an overwhelmingly large
component from the normal distribution with mean corresponding to
large dissipation, with a negligible contribution arising from a rare
non dissipative event corresponding to the refolding in the
$\alpha$-helix structure.  For short and intermediate duration
times, the ``refolding'' of $A_{10}$ has an high chance to fail
producing a manifold of misfolded structures. As the duration time
grows, the likelihood of the non dissipative process (i.e. the correct
refolding) grows as well. The break-even point for the likelihood of
two events in the reverse driven process occurs between duration times
of 0.2 and 0.3 ns.  The above results suggests a possible route in
real experiments on single molecules using, e.g., an optical trap
apparatus to detect metastable states.  In fast pulling experiments,
the extra energy implied in the large dissipation allows to visit
states that are hard to visit in a driven quasi reversible
experiment. In presence of two competing minima, one could then use
the dual Markov model extrapolated from few bidirectional work
measurements to both achieve, trough the KL divergence, and its
connection to the dissipation, a better estimate of the free energy
between the final and initial states and to identify secondary
metastable minima at fixed driven coordinate that are difficult to
evaluate either because of the presence of high barrier or because
they are several KT larger than the principal (native) structure.
\newpage
\bibliography{ms}
  
\clearpage
\begin {center}
{\bf \large Tables} 
\end{center}
\begin{table}[b!]
\begin{tabular}{cccccc}
\hline
$~~~~~~~~~\tau~~~~~~~$ & $\Delta F$ & $<W_f^d>$ & $\sigma^2_f$ & $<W_r^d>$ & $\sigma^2_r$ \\ 
\hline
     0.021 &   86.0 & 83.2   & 245.6  & 82.8 &   67.7 \\
     0.042 &   84.2 & 60.4   & 174.6  & 72.9 &  142.0 \\
     0.063 &   94.0 & 40.6   & 133.4  & 75.2 &  282.4 \\
     0.084 &   92.3 & 34.7   & 114.0  & 65.3 &  406.2 \\
     0.105 &   93.8 & 28.6   & 105.8  & 59.5 &  489.5 \\
     0.150 &   93.9 & 22.8   &  75.6  & 46.6 &  507.8 \\
     0.210 &   93.7 & 18.6   &  69.4  & 35.6 &  434.0 \\
     0.300 &   93.9 & 14.1   &  54.4  & 25.6 &  303.9 \\
     0.420 &   92.9 & 12.0   &  52.1  & 18.3 &  204.4 \\
     0.630 &   93.0 &  8.7   &  42.8  & 11.9 &  105.9 \\
     0.840 &   93.2 &  7.1   &  32.6  &  9.2 &   79.4 \\
     0.930 &   93.6 &  6.3   &  30.1  &  8.4 &   88.3 \\
     1.050 &   93.2 &  5.8   &  24.8  &  7.9 &   85.5 \\
     2.100 &   93.4 &  3.2   &  16.3  &  3.9 &   40.4 \\
     4.200 &   93.2 &  1.7   &   9.7  &  2.3 &   33.4 \\
\hline
\end{tabular}
\caption{\label{table1}Salient data of the work distributions in
  Alanine deca-peptide {\it in  vacuo} at 300 K. For each duration
  time $\tau$, the forward and reverse work distributions have been
  calculated using 504  trajectories. $\Delta F$ is the free energy
difference between the final (all trans extended structure,
$\zeta=31.5$ \AA)   and the initial ($\alpha$-helix structure, 
$\zeta=15.5$ \AA) using the Bennett acceptance ratio\cite{shirts03}
on the 1008 forward and reverse trajectories. $<W_f^d>$, $\sigma_f^2$  
$<W_r^d>$, $\sigma_r^2$ are the mean dissipated work and variance of the forward and
reverse work distributions, respectively}
\end{table}
\clearpage
\begin{table}[b!]
\begin{tabular}{ c c c c c c c c c c c c c}
\hline \hline
& & \multicolumn{5}{c}{Forward distribution} & \multicolumn{5}{c}{Reverse distribution} \\
\cline{3-7} \cline{9-13}
$\tau$ & KL & $p$ & $w_1$ & $\sigma_1$  &  $w_2$ & $\sigma_2$ &  & $q$
& $w_3$ & $\sigma_3$  &  $w_4$ & $\sigma_4$ \\  
\hline
0.021  &   0.38  &   1.00  & 167.38  & 314.83  &  21.44  &  47.49 &~~~~~&    0.99  &  2.40  & 47.49 &  41.15 & 314.83 \\
0.042  &   0.55  &   1.00  & 143.35  & 222.69  &  31.04  &  60.63 &~~~~~&    0.90  &  6.73  & 60.63 &  54.06 & 222.69 \\
0.063  &   0.48  &   1.00  & 134.87  & 187.48  &  41.19  &  72.37 &~~~~~&    0.82  & 12.17  & 72.37 &  59.69 & 187.48 \\
0.084  &   0.60  &   1.00  & 126.43  & 144.10  &  99.73  & 199.01 &~~~~~&    0.84  & 19.94  &199.01 &  68.65 & 144.10 \\
0.105  &   0.43  &   1.00  & 122.30  & 126.78  & 118.34  & 231.15 &~~~~~&    0.79  & 25.65  &231.15 &  71.47 & 126.78 \\
0.150  &   0.51  &   0.99  & 115.49  &  97.68  & 131.40  & 240.41 &~~~~~&    0.69  & 35.00  &240.41 &  76.32 &  97.68 \\
0.210  &   0.42  &   0.98  & 111.76  &  82.83  & 124.30  & 199.57 &~~~~~&    0.59  & 44.28  &199.57 &  78.55 &  82.83 \\
0.300  &   0.33  &   0.97  & 107.78  &  65.06  & 115.62  & 149.68 &~~~~~&    0.51  & 55.60  &149.68 &  81.69 &  65.06 \\
0.420  &   0.41  &   0.97  & 104.62  &  52.95  & 113.55  & 133.20 &~~~~~&    0.65  & 83.39  & 52.95 &  60.14 & 133.20 \\
0.630  &   0.31  &   0.99  & 102.49  &  44.96  & 106.00  &  97.50 &~~~~~&    0.82  & 84.46  & 44.96 &  66.91 &  97.50 \\
0.840  &   0.37  &   1.00  & 102.07  &  44.96  & 113.08  & 148.18 &~~~~~&    0.98  & 84.04  & 44.96 &  53.67 & 148.18 \\
0.930  &   0.27  &   1.00  & 100.17  &  35.42  &  99.68  & 123.54 &~~~~~&    0.97  & 85.97  & 35.42 &  50.15 & 123.54 \\
1.050  &   0.31  &   1.00  &  99.40  &  31.53  & 102.81  & 124.17 &~~~~~&    0.97  & 86.76  & 31.53 &  53.02 & 124.17 \\
2.100  &   0.27  &   0.99  &  96.26  &  16.09  & 115.83  & 123.66 &~~~~~&    0.98  & 89.81  & 16.09 &  66.25 & 123.66 \\
4.200  &   0.14  &   0.89  &  94.56  &   7.41  &  96.67  &  21.06 &~~~~~&    0.86  & 91.59  &  7.41 &  88.23 &  21.06 \\
\hline
\end{tabular}
\caption{\label{table2}Best fit parameters for the true forward and reverse
  distributions, using a bimodal distribution Eq. \ref{eq:dual}. $\tau$ is the duration time of the non equilibrium
  experiment in ns. $KL$ is the Kullback-Leibler divergence (in kJ
  units) between
  the sum of the true forward and reverse distribution and the sum of the fitted bimodal
  forward and reverse distributions. $w_3\equiv w_1 - \beta \sigma_1^2$ and $w_4\equiv w_2
  - \beta \sigma_2^2$ are the mean value of the normal distribution of
  the reverse process.  For the meaning of the other symbol see
  text. Units of energy are kJ mol$^{-1}$.  }
\end{table}
\clearpage
\begin{table}[b!]
\begin{tabular}{ c c c c }
\hline 
\hline 
$\tau$ & KL & $w$ & $\sigma$  \\
\hline 

0.021 & 2.20 &   177.6  &  421.811  \\ 
0.042 & 5.14 &   176.6  &  416.912  \\ 
0.063 & 6.16 &   172.2  &  395.072  \\ 
0.084 & 7.25 &   130.5  &  187.082  \\ 
0.105 & 7.06 &   125.9  &  164.197  \\ 
0.150 & 7.98 &   117.7  &  123.593  \\ 
0.210 & 5.13 &   114.1  &  105.130  \\ 
0.300 & 3.23 &   109.5  &   81.816  \\ 
0.420 & 3.01 &   105.6  &   62.393  \\ 
0.630 & 1.44 &   102.2  &   45.533  \\ 
0.840 & 1.01 &   100.3  &   36.747  \\ 
0.930 & 1.95 &    98.9  &   29.519  \\ 
1.050 & 1.80 &    99.4  &   31.581  \\ 
2.100 & 1.92 &    96.1  &   15.379  \\ 
4.200 & 0.32 &    94.7  &    8.348  \\
\hline 
\end{tabular}
\caption{\label{table3}Best fit parameters of the true forward and reverse
  distribution using the linear model. $KL$(kJ mol$^-1$)  is the Kullback-Leibler divergence between
  the sum of the true forward and reverse distributions and sum of the fitted Gaussian
  forward and reverse  distributions.}
\end{table}
\clearpage

{\bf \large Figure Captions}

\begin{list}{}{\leftmargin 2cm \labelwidth 1.5cm \labelsep 0.5cm}

\item[{\bf Fig. 1}] Forward (on the right in brown) and backward (on
  the left in black) work distributions for Alanine decapeptide {\it
    in vacuo} at 300 K obtained in non equilibrium experiments of
  various duration time ranging, from bottom to top, from 0.021 s to
  4.2 ns. Each forward and reverse distribution has been calculated
  using 504 work measurements.

\item[{\bf Fig. 2}] Mean dissipation {\it vs} the duration time of 
the non equilibrium experiments for the forward (unfolding) and
reverse (refolding)  of Alanine decapeptide {\it vacuo}. 

\item[{\bf Fig. 3}] Jensen-Shannon divergence {\it vs} the mean
  dissipation $0.5 (<W_f> + <W_r>)$ in Alanine decapeptide {\it vacuo}
  at 300 K.  The triangle symbols have been calculated according to
  Eq.  \ref{eq:js}. The circle have calculated using Eq. 7 of Ref.
  \onlinecite{feng08} by using a $\Delta F$ of 93.3 kJ mol$^{-1}$.
  The solid line refers to a Gaussian (Markovian) model such that
  $<W_d> = \beta \sigma^2/2$ in both forward and reverse directions.

\item[{\bf Fig. 4}] {\bf (a)}: Forward (circle and solid line) and reverse
  (triangle and dotted line) distribution for a duration time of
  $\tau=0.105$ ns in $A_{10}$ {\it in vacuo} at 300 K compared with the
  best fit Markov model (solid and dashed thick lines ) satisfying
  Eq. \ref{eq:crooksw}. 

  {\bf (b)}: Same as in a) except for a duration time of
  $\tau=0.210$ ns.

\item[{\bf Fig. 5}] Probabilities $p$ and $q$ (see
  Eq. \ref{eq:ctbimod}) for a bimodal distribution satisfying the CT 
(Eq. \ref{eq:crooksw})  
  as a function of $x = \Delta F-w_1 + \frac{1}{2} \beta\sigma_1^2$ and 
$y = \Delta F-w_2 + \frac{1}{2} \beta\sigma_2^2$ ($RT$ units).  

\item[{\bf Fig. 6}] True and fitted forward and reverse work
  distributions in Alanine decapeptide {\it in vacuo} at 300 K for 
  various duration times of the non equilibrium experiments using
the bimodal approach, Eq. \ref{eq:dual}. 
The forward and reverse true distribution are in brown and black,
respectively. 
The forward and reverse fitted distribution are in violet  and blue,
respectively.  The parameters of the fit can be found in Table \ref{table2}. 

\end{list}

\clearpage
\newpage

\clearpage
\newpage

\clearpage
\newpage

\clearpage
\newpage
\begin{figure}
\begin{center}
\caption{} 
\label{fig4}
\includegraphics[scale=0.8,clip]{FIGS/fig4.eps}
\end{center}
\end{figure}

\clearpage
\newpage
\begin{figure}
\begin{center}
\caption{} 
\label{fig5}
\includegraphics[scale=0.6,angle=-90,clip]{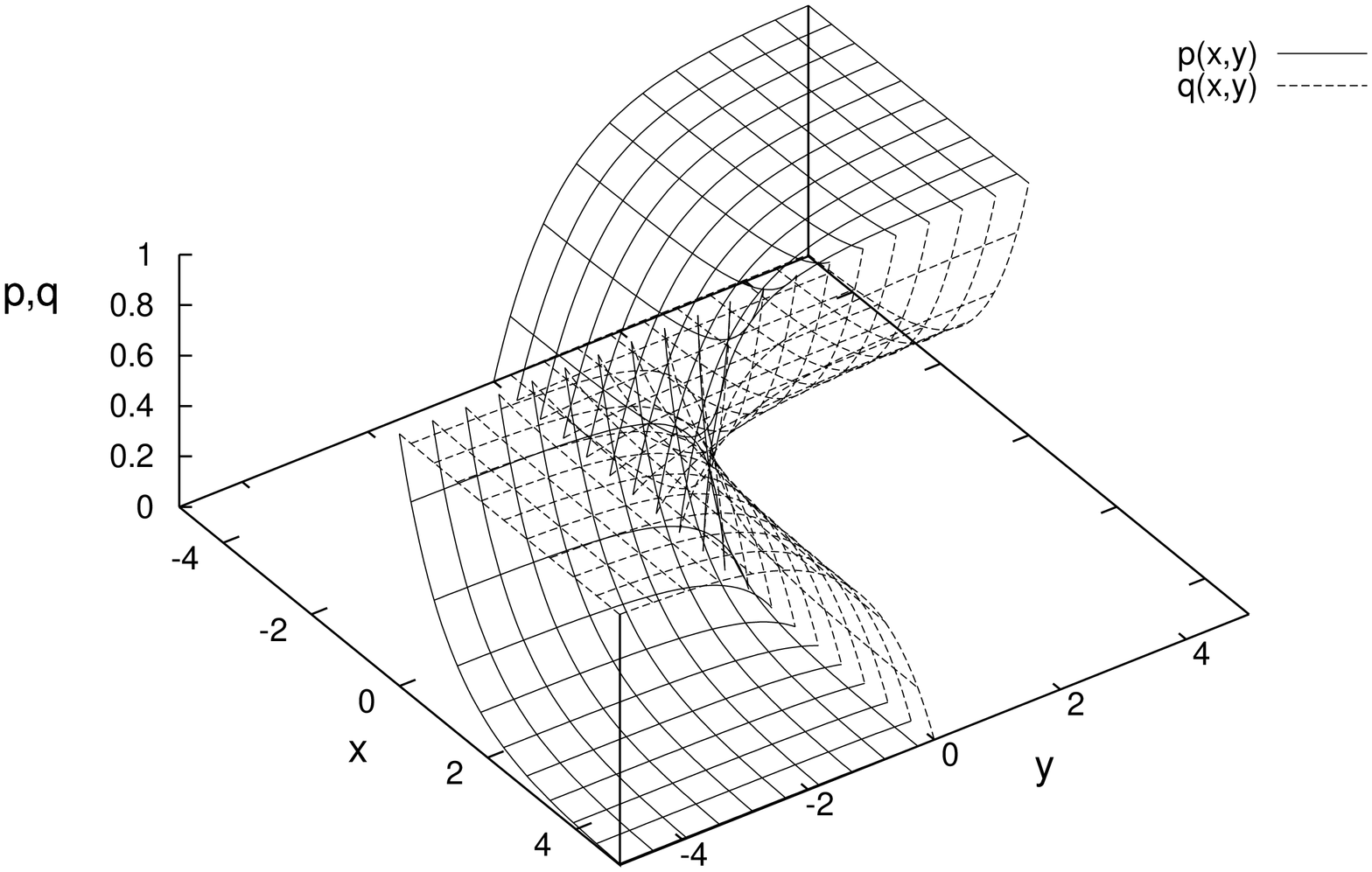}
\end{center}
\end{figure}

\clearpage
\newpage
\begin{figure}
\begin{center}
\caption{} 
\label{fig6}
\includegraphics[scale=0.7,clip]{FIGS/fig6.eps}
\end{center}
\end{figure}

\end{document}